\documentclass[12pt]{article}

\catcode`\@=11
\global\arraycolsep=2pt
\usepackage{amsbsy,amssymb,latexsym,amsfonts,amsmath}
\usepackage{graphicx,color}

\begin{document}
\begin{flushright}
\parbox{4.2cm}
{RUP-19-19}
\end{flushright}

\vspace*{0.7cm}

\begin{center}
{ \Large Conformal Contact Terms and Semi-Local Terms}
\vspace*{1.5cm}\\
{Yu Nakayama}
\end{center}
\vspace*{1.0cm}
\begin{center}

Department of Physics, Rikkyo University, Toshima, Tokyo 171-8501, Japan

\vspace{3.8cm}
\end{center}

\begin{abstract}
We study conformal properties of local terms such as contact terms and semi-local terms in correlation functions of a conformal field theory. Not all of them are universal observables but they do appear in physically important correlation functions such as (anomalous) Ward-Takahashi identities or Schwinger-Dynson equations. We develop some tools such as embedding space delta functions and effective action to examine  conformal invariance of these local terms.
\end{abstract}

\thispagestyle{empty} 

\setcounter{page}{0}

\newpage


%-----------------------------------------

%\pacs{}
%\keywords{Suggested keywords}%Use shokeys class option if keyword
%\maketitle

%%%%%%%%%%%%%%%%%%%%%%%%%%%%%%%%%%%%%%%%%%%%%%%%%%%%
\section{Introduction}
Correlation functions of local operators are fundamental observables in conformal field theories. They are non-local functions of space and they possess severe constraints from the conformal symmetry and the operator product expansions. These constraints are the basis of the conformal bootstrap approach to conformal field theories, which played a pivotal role in our current understanding of critical phenomena in higher dimensions than two. See e.g. \cite{Poland:2018epd} for a recent review.

They are relatively less studied but local terms in the correlation functions are sometimes physically important. By ``local" we mean they have  at least one delta function (appropriately regularized) in the coordinate space correlation functions. When the support of the correlation functions is just one point, we may call them ``contact terms". Otherwise we may call them ``semi-local". 

To list a few physically relevant local terms, we have canonical commutation relations, Schwinger-Dyson equations, and (anomalous) Ward-Takahashi identities. It is noted that they are more or less related to the ``Lagrangian picture" of the quantum field theory. It is therefore a good meeting point to understand the role of local terms in axiomatized conformal field theories, which tries to dismiss the smell of ``Lagrangian".

More recently, we have learned that some local terms play important roles to understand the renormalization group flow and the moduli space of coupling constants. The  effects of local terms have played crucial roles in the momentum space correlation functions in particular to understand their ultraviolet behavior. For example, the dilaton scattering amplitudes \cite{Komargodski:2011vj} that are used to understand the renormalization group flow may be affected by the semi-local terms \cite{Bzowski:2014qja}\cite{Dymarsky:2014zja}\cite{Bzowski:2017poo}. The semi-local terms are also important in our understanding of shortening anomalies or anomalies in uplifting coupling constants to background fields \cite{Gomis:2015yaa}\cite{Gomis:2016sab}\cite{Schwimmer:2018hdl}\cite{Cordova:2019jnf}. 

Local terms in correlation functions are delicate objects because they are often affected by local counter-terms as well as a definition of the operator itself. Some local terms, however, have no such ambiguities from the symmetry consideration and they are universal (see some examples in \cite{Maldacena:2011nz}\cite{Closset:2012vg}\cite{Closset:2012vp}). 
The lack of universality does not necessarily mean that they are physically irrelevant. They are important, similarly to the cosmological constant, as a part of the physical model, but they are simply independent of the framework of the conformal field theories and we have less predictive powers.

In this paper, we develop some techniques to study local terms in conformal field theories. Along the lines, we study examples of conformal invariant contact terms or semi-local terms in physically relevant correlation functions. In section 2, we introduce representations of a coordinate-space delta function in the embedding space formalism. In section 3, we study contact terms in two-point functions. In section 4 we study local terms in three-point functions. In section 5 we conclude with further discussions.

\section{Embedding space delta functions}
In this paper, we study conformal contact terms and semi-local terms in correlation functions of local operators. By definition, these correlation functions include at least one delta function in coordinate space. It is therefore important to understand the conformal properties of the delta function in the coordinate space. While we may study these properties in the coordinate space directly (see Appendix A for details), it is often convenient to use the embedding space formalism \cite{Costa:2011mg}, making the obscured conformal symmetry manifest as a $D=d+2$ dimensional Lorentz symmetry. Our first goal, therefore, is to establish  representations of a delta function in the embedding space formalism.\footnote{Conformal invariant integration in the embedding space formalism was studied in \cite{SimmonsDuffin:2012uy}.}

To fix the notation, we use the $D=d+2$ dimensional embedding space coordinate $X^A$ by extending the $d$-dimensional coordinate $x^\mu$ as
\begin{align}
X^A = (X^+, X^-,X^\mu)
\end{align}
and define the $d+2$ dimensional metric
\begin{align}
ds^2 &= \eta_{AB} dX^A dX^B \cr
& = -dX^+ dX^- + \delta_{\mu\nu} dX^\mu dX^\nu \ . 
\end{align}
Throughout the paper, we work on Euclidean conformal field theories, so the $d$ dimensional metric is given by the Euclidean Kronecker delta $\delta_{\mu\nu}$.
The physical fields are defined on the light-cone
\begin{align}
X^2 = \eta_{AB} X^A X^B = 0
\end{align}
and we further introduce the projective identification $X^A = \lambda X^A$. These conditions are $SO(1,d+1)$ invariant, and this $D$-dimensional Lorentz symmetry will be identified with the $d$-dimensional conformal symmetry. 

A particular section of the light-cone
\begin{align}
X^A = (1,x^2,x^\mu)
\end{align}
is called the Poincar\'e section. This section is suitable for the study of the conformal correlation functions on $d$-dimensional flat Euclidean space.

Conformal primary fields in $d$-dimension are uplifted to the $D$-dimensional fields $\Phi_{ABC\cdots}(X^A)$ with the following properties
\begin{itemize}
\item
 They are defined on the light-cone $X^2 = 0$.
\item
They are degree $-\Delta_\Phi$ functions of projective coordinates i.e. $\Phi_{AB\cdots}(\lambda X) = \lambda^{-\Delta_\Phi} \Phi_{AB\cdots}(X)$. Here the minus of the degree (i.e. $\Delta_{\phi}$) will be identified with the scaling dimension, which we also call the projective weight. 
\item
They are transverse: $X^A \Phi_{ABC\cdots}(X^A) =  X^B \Phi_{ABC\cdots}(X^A)  = \cdots =0$.
\end{itemize}
We are mostly working on symmetric traceless tensors, but the other representations have been studied in the literature \cite{Kravchuk:2016qvl}\cite{Cuomo:2017wme}\cite{Isono:2017grm}\cite{Karateev:2019pvw}\cite{Fortin:2019dnq}.

Because of these properties, the basic  $SO(1,d+1)$ invariants are scalar homogeneous functions $X_i X_j (= \eta_{AB} X^A_i X^B_j)$ and transverse tensor structures such as $\eta_{MN} -\frac{X_{jM} X_{iN}}{X_i \cdot X_j}$. In order to study contact terms and semi-local correlation functions, however, we would like to represent a delta function in the embedding space language. The $D$-dimensional delta function $\delta^{(D)}(X-Y)$ is obviously $SO(1,d+1)$ invariant, but this is not well-defined on the projective light-cone, so we need  more elaborate distributions.

We introduce the embedding-space delta functions by
\begin{align}
\delta^{(d)}_{k}(X,Y) = \int_{-\infty}^{\infty} \frac{ds}{s^{k+1}} \left( \int_{-\infty}^{\infty} d(R^2) \delta^{(D)} (X-sY) \right)|_{X^2 = Y^2 = 0}  \ . \label{emdelta}
\end{align}
Here $R^2 = \eta_{MN} X^M X^N$ (defined outside of the light-cone as well), and we set $X^2 = Y^2=0$ after the integration over $R^2$. The integration over $s$ guarantees that the left hand side has the support when $X^A = s Y^A$ for a particular $s$, which is precisely the condition that the two-points represented by $X^A$ and $Y^A$ are identical on the projective light-cone.\footnote{The introduction of $s$ integration to represent the delta functions on the projective space can be found in the twistor literature (e.g. \cite{Mason:2011nw}). We need an extra integration over $R^2$ to restrict the delta functions on the light-cone.}

Now we present some properties of these delta-functions. First of all, they are Lorentz invariant in $D$-dimensions and defined on the light cones $X^2 = Y^2=0$. While they all give the same (up to possible normalization constants) delta function in the Poincar\'e section, they have apparent one-parameter dependence on $k$. This $k$ dependence determines the projective weights of the delta functions as we will see. 

Under the scaling of the first argument $X^A \to \lambda X^A$, the embedding space delta functions transform as $\delta^{(d)}_{k}(\lambda X,Y) = \lambda^{2-D-k} \delta^{(d)}_{k}(X,Y)$. Similarly, under the scaling of the second argument $Y^A \to \lambda Y^A$, they transform as  $\delta^{(d)}_{k}(X, \lambda Y) \to \lambda^{k} \delta^{(d)}_{k}(X,Y)$. Thus, the embedding space delta functions have $k$ dependent projective weights $(-(2-D-k),-k)$ for $X^A$ and $Y^A$ respectively. We will see that it plays an important role that we can adjust the (relative) weights by choosing $k$ differently while it gives the same delta function in the Poincar\'e section. Note, however, that the sum of the projective weights for $X^A$ and $Y^A$ is always given by $D-2=d$, which is related to the fact that the $d$-dimensional delta function $\delta^{(d)}(x-y)$ has the unique scaling dimension $d$ as $(x-y)^{-d}$ in the Poincar\'e section.

A technical comment is in order. In our definition of the embedding space delta functions \eqref{emdelta}, the integration measure $d(R^2)$ looks arbitrary. Indeed, one can alternatively use $d(\tilde R^2)$, where $\tilde{R}^2 = \eta_{MN} Y^M Y_N$, resulting in different projective weights in $X^A$ and $Y^A$. However, it is not possible to change the total weight (near the origin $R^2 =0$ which is only relevant for us) by multiplying further powers of $X^2$, $Y^2$ or $XY$ inside the integration. Since we will eventually evaluate them on the light-cone $X^2 = Y^2$ (and consequently $XY$=0 from the delta function), the additional multiplication makes them either vanish or infinite (as a distribution), so for our purpose of representing the coordinate space delta function in the embedding space formalism, we do not need them.\footnote{More precisely, with more careful distributional analysis, it is not impossible to make sense of such singular terms as $ \delta(x) \mathrm{P}\frac{1}{x} = -\frac{1}{2}\delta'(x) $, where $\mathrm{P}$ represents the principal value, a la Sato \cite{Sato}, but they typically fail to be associative. In this case, the products of $x$, $\delta(x)$ and $\mathrm{P}\frac{1}{x}$ can be $0$, $\frac{1}{2}\delta(x)$, $\delta(x)$ depending on the order. We will not study such terms in this paper.}

Note that  it is a special feature of embedding space delta functions that the (relative) projective weights can be changed at will. We will later see that the derivatives of embedding space delta functions do not have such a property, which is deeply connected with the representation theory of the conformal algebra.

\section{Two-point functions}
Let us first consider the simplest example of conformal contact terms given by two-point functions of scalar primary operators. By definition, there are no semi-local terms in two-point functions. Using the embedding space formalism and the delta functions defined in section 2, we can immediately write down the candidate
\begin{align}
\langle \Phi_1 (X_1) \Phi_2(X_2) \rangle = c_{12} \delta^{(d)}_k(X_1,X_2) \ .
\end{align}
The projective weight must satisfy $\Delta_1 = d+k$ and $\Delta_2 = -k$ by comparing the weights in the both sides. 

In the Poincar\'e section, it reduces to the contact term 
\begin{align}
\langle \Phi_1 (x_1) \Phi_2(x_2) \rangle = c_{12} \delta^{(d)}(x_1-x_2) \ . 
\end{align}
Our embedding space formalism implies that this contact term is conformal invariant as long as $\Delta_1 + \Delta_2 = d$. This is in sharp contrast with the conformal invariant non-local two-point functions in which the conformal invariance dictates that the conformal weights must be identical $\Delta_1 = \Delta_2$ to obtain a  non-zero result.

We can directly check that the coordinate space delta function is conformal invariant as long as $\Delta_1 + \Delta_2 = d$, but we may alternatively study the effective functional. Suppose correlation functions are defined by deriving the partition functional $Z[J] = \int d\Phi e^{-S_{\mathrm{eff}}[J]}$ with respect to the source $J^i$ introduced in the action:
\begin{align}
S_{\mathrm{eff}} = S_0 + \int d^d x \sum_i J^i(x) \Phi_i(x) \ .
\end{align}
Let us now try to add the local term $\int d^dx c_{12} J^1(x) J^2(x)$ to the effective action. Then the two-point function acquires the additional contact term
\begin{align}
\langle \Phi_1(x_1) \Phi_2(x_2) \rangle = \frac{\delta^2 Z[J]}{\delta J^1(x_1) \delta J^2(x_2)}  =  \langle \Phi_1(x_1) \Phi_2(x_2) \rangle_{c=0} + c_{12} \delta^{(d)}(x_1-x_2) \ .
\end{align}
Now the added term in the action is (spuriously) Weyl invariant as long as $\Delta_{J^1} + \Delta_{J^2} = d$ or equivalently $\Delta_{\Phi_1} + \Delta_{\Phi_2} = d$ (because $\Delta_{J^i} + \Delta_{\Phi_i} = d$). This indicates that the coordinate space delta function is conformal invariant as long as $\Delta_{1} + \Delta_{2} = d$. The condition is weaker than $\Delta_{1} = \Delta_{2}$, which can be seen that in the above effective action approach, the non-local part of $\langle \Phi_1(x_1) \Phi_2(x_2) \rangle_{c=0} =0$ (when $\Delta_1 \neq \Delta_2$).

It is immediate to generalize the construction above  with the spinning two-point functions as long as we have just a delta function on the right hand side. For example, the two-point functions of vector primary operators may include the contact term
\begin{align}
\langle \Phi_1^\mu(x_1) \Phi_2^\nu(x_2) \rangle = c_{12} \delta^{\mu\nu} \delta^{(d)}(x_1-x_2)
\end{align}
as long as $\Delta_1 + \Delta_2 = d$. The corresponding effective action is $\int d^dx J^\mu(x) J_\mu(x)$. 
The most famous example would be $\langle \partial^\mu \phi(x_1) \partial^\nu \phi (x_2) \rangle$ of a free scalar in $d=2$ dimensions.

Now let us consider the case when the contact term is given by derivatives of a delta function. The distinctive point here is that unlike contact terms without derivatives discussed above, the projective weights are completely fixed. One may naively want to use the embedding space derivatives $\frac{\partial}{\partial X^M}$ acting on the delta functions $\delta^{(d)}_{k}(X,Y)$, but as emphasized in \cite{Costa:2011mg}, the derivative is generically not well-defined as an operator acting on functions on the projective light-cone. Only when they have a certain projective weight, the (specific combinations of) derivatives can be properly defined.

Let us study the simplest example of $\partial_\mu \delta^{(d)}(x_1-x_2)$. It should be related to $\partial_M^1 = \frac{\partial}{\partial X_1^M}$ acting on the embedding space delta functions  $\delta^{(d)}_{k}(X_1,X_2)$. Considering the transverse condition as well, we see that the projective weight on $X_1$ must be zero. In other words,
\begin{align}
\langle \Phi_{M1} (X_1) \Phi_2(X_2) \rangle = \partial^{1}_M \delta^{(d)}_k(X_1,X_2)
\end{align}
is well-defined only when $\Delta_1 = 1$ and $\Delta_2 = d$ so that $k=-d$. In the Poincar\'e section, it is given by
\begin{align}
\langle \Phi_{\mu1} (x) \Phi_2(y) \rangle = \partial_\mu \delta^{(d)}(x-y) \ .
\end{align}

By using the effective action, we can understand why this is the only allowed conformal dimensions. The effective action that reproduces the delta function with one derivative is 
\begin{align}
S_{\mathrm{eff}} = \int d^d x \partial^\mu J_\mu^1(x) J^2(x) \ .
\end{align}
However, one can convince oneself that this effective action is conformal invariant (or Weyl invariant) only when $\Delta_{J^1_\mu} = d-1$ and $\Delta_{J^2} = 0$. This can be understood by observing that the conservation of a vector primary operator is Weyl invariant only if $\Delta_{J_\mu} = d-1$ (or equivalently a derivative of a scalar primary is again a primary operator only when it has $\Delta_{J}=0$).
 Therefore the two-point function is not conformal invariant unless $\Delta_{1} = 1$ for $\Phi_{\mu 1}$ and $\Delta_{2}=d$ for $\Phi_2$. This condition is completely dictated by the representation theory of the conformal algebra, and the embedding space formalism gives a further verification.

The next non-trivial example is the Laplacian of the delta function. By using the technique developed in \cite{Costa:2011mg}, we can show
\begin{align}
\delta^{\mu\nu} \partial_\mu \partial_\nu \delta^{(d)}(x_1-x_2) = \left( \eta^{MN} \partial^{1}_M \partial^1_N + (d -2(\Delta+1))\bar{X}^A\partial_A^{1}  \right) \delta^{(d)}_{k}(X_1,X_2) \ 
\end{align}
with $\bar{X}^A = (0,2,\vec{0})$
in the Poincar\'e section, so the Laplacian acting on the delta function is conformal invariant only when the projective weight $\Delta$ of the embedding space delta function on the right hand side (for the first argument $X_1$) is $ \frac{d}{2}-1$ or $k=-\frac{d}{2}-1$. This is also expected from the representation theory of the conformal algebra, which says that the Laplacian is conformal invariant (only) when it acts on a scalar operator with dimension $\Delta = \frac{d}{2}-1$.
Therefore, the conformal two-point functions may contain the Laplacian of the delta function 
\begin{align}
\langle \Phi_1(x_1) \Phi_2(x_2) \rangle = c_{12} \partial^\mu \partial_\mu \delta^{(d)}(x_1-x_2)
\end{align}
when $\Delta_1 = \frac{d}{2}+1$ and $\Delta_2 = \frac{d}{2}+1$.

Again, one can come up with the simple effective action 
\begin{align}
S_{\mathrm{eff}} =  \int d^dx J^1(x) \partial^\mu \partial_\mu J^2(x) \ ,
\end{align}
which is conformal invariant when $\Delta_{J_1} = \Delta_{J_2} = \frac{d}{2}-1$. This is well-known from the fact that the the free scalar is conformal invariant only if we assign the definite scaling dimensions $\Delta = \frac{d}{2}-1$.

One can keep going and study exactly when the derivatives of delta functions are conformal invariant in our formulation, but presumably it is better not to reinvent the wheel but to just borrow the known results in  the representation theory of conformal algebra (see \cite{Penedones:2015aga} and reference therein), which tells us when and how the differential operators become null (meaning conformal invariant). Indeed all the null vectors can be represented in the effective action in this way and so are the conformal invariant derivatives of the delta function.

Let us discuss how the conformal contact terms appear physically in known two-point functions. The most common case comes from the conformal anomaly. The conformal anomaly of two-point functions mean that if we change the renormalization parameter such as a cut-off, the two-point functions acquire the additional delta function (while the non-local part is completely fixed by the conformal symmetry). For scalar primary operators, the simplest case happens when $\Delta_\Phi = \frac{d}{2}$: under the scale transformation $\delta_{\sigma}$, 
\begin{align}
\delta_{\sigma} \langle \Phi(x_1) \Phi(x_2) \rangle  = \delta_{\sigma} \left( \frac{1}{(x_1-x_2)^d}\right)_{\mathrm{reg}} = \sigma \delta^{(d)}(x_1-x_2) \ . \label{shift}
\end{align}
Note that we need a certain regularization to give a precise meaning to the naive conformal invariant expression $\frac{1}{(x_1-x_2)^d}$ as a distribution. For example, in the momentum space, the Fourier transform is logarithmically divergent and it has a cut-off dependence as $\log\frac{k^2}{\sigma^2}$. When the cut-off is changed, the local term is shifted as in \eqref{shift}.

Similarly, delta functions with derivatives can occur in scalar primary two-point functions as a part of the trace anomaly when $2 \Delta_{\Phi} = d + 2n$. Examples are (exactly) marginal operators in $d=4$ dimensions with $ (\partial^\mu \partial_\mu)^2 \delta^{(d)}(x_1-x_2)$ or in $d=2$ dimensions with  $ (\partial^\mu \partial_\mu) \delta^{(d)}(x_1-x_2)$.\footnote{An interesting question arises if  these conformal invariant contact terms can be compatible with the Weyl invariance. \cite{Schwimmer:2019efk}\cite{Nakayama:2019xzz} showed that it is not necessarily possible with higher orders of derivatives as one can see that the higher powers of Laplacians are not always Weyl invariant (in even dimensions).}

So far, our examples have $\Delta_1 = \Delta_2$ so that the non-local two-point functions are also possible. Now let us consider the more exotic case with $\Delta_1 \neq \Delta_2$, in which the non-local two-point functions do not exist. In a terminology of \cite{Nakayama:2018dig}, it is related to an ``impossible anomaly". The non-existence of the corresponding  non-local terms make these correlation functions very distinct.

The simplest such an example is a free scalar two-point function. It has the non-local form of 
\begin{align}
\langle \phi(x_1) \phi (x_2) \rangle = \frac{1}{(x_1-x_2)^{d-2}} .
\end{align}
Acting the Laplacian, it shows a conformal invariant contact term.
\begin{align}
\langle \partial^\mu \partial_\mu \phi(x_1) \phi(x_2) \rangle = c_d \delta^{(d)}(x_1-x_2) 
\end{align}
with a definite coefficient $c_d$. Note that the representation theory of conformal algebra tells that $\partial^\mu \partial_\mu \phi(x)$ corresponds to a null state so it transforms as a primary operator. In this way, we have constructed an example of conformal invariant two-point functions with different conformal dimensions (i.e. $\Delta_1 = \frac{d}{2}+1$ and $\Delta_2 = \frac{d}{2}-1$) as a contact term.

Note that in this case, the coefficient $c_d$ has a definite meaning associated with the Schwinger-Dyson equations of a free scalar. While we treat as if $\partial^\mu \partial_\mu \phi$ and $\phi$ were independent primary operators when we study the conformal properties of the contact term, they are not, and we cannot introduce the local counter-term that changes the coefficient $c_d$ as would be possible if they were independent.\footnote{In the Hamiltonian picture, this delta function originates from the $T$-product and the canonical commutation relation.}

\section{Three-point functions and more}
We next consider three-point functions of conformal primary operators. In addition to completely local  contact terms with two delta functions, semi-local terms with only one delta function are possible. 

The simplest example is a contact term with three scalar primary operators. Without using the derivative of delta functions, we can construct conformal invariant three-point functions by using the embedding space formalism as
\begin{align}
\langle \Phi_1(X_1) \Phi_2 (X_2) \Phi_3(X_3) \rangle = \delta^{(d)}_{k}(X_1,X_2) \delta^{(d)}_{\tilde{k}}(X_2,X_3) \ ,
\end{align}
where $\Delta_1 = d+k$, $\Delta_2 = -k+d+\tilde{k}$, $\Delta_3 = -\tilde{k}$. In the Poincar\'e section, it is evaluated as
\begin{align}
\langle \Phi_1(x_1) \Phi_2 (x_2) \Phi_3(x_3) \rangle = \delta^{(d)}(x_1-x_2)\delta^{(d)}(x_2-x_3) \ \label{3cont}
\end{align}
with $\Delta_{1} + \Delta_2 + \Delta_3 = 2d$. 

One could have used the other expressions such as
\begin{align}
\langle \Phi_1(X_1) \Phi_2 (X_2) \Phi_3(X_3) \rangle = \delta^{(d)}_{k'}(X_1,X_3) \delta^{(d)}_{\tilde{k}'}(X_2,X_3) \ ,
\end{align}
where $\Delta_1 = d+k'$, $\Delta_2 = d+\tilde{k}'  $, $\Delta_3 = -k' -\tilde{k}'$, which give the same distribution as \eqref{3cont} in the Poincar\'e section. The corresponding effective action is 
\begin{align}
S_{\mathrm{eff}} = \int d^d x J_1(x) J_2(x) J_3(x) \ , 
\end{align}
which is conformal invariant if $\Delta_{J_1} + \Delta_{J_2} + \Delta_{J_3} = d$ or $\Delta_{1} + \Delta_2 + \Delta_3 = 2d$ as expected. Note that unlike the two-point contact terms, all of them here may have corresponding non-local three-point functions.

Similarly one can construct conformal invariant semi-local scalar three-point functions from the embedding space formalism.
\begin{align}
\langle \Phi_1(X_1) \Phi_2 (X_2) \Phi_3(X_3) \rangle = \delta^{(d)}_{k}(X_1,X_2) (X_2 X_3)^{-\delta}(X_1 X_3)^{-\tilde{\delta}} \ , 
\end{align}
where $\Delta_1 = d+k + \tilde{\delta}$, $\Delta_2 = -k+\delta  $, $\Delta_3 = \delta + \tilde{\delta} $. In the Poincar\'e section, it is evaluated as
\begin{align}
\langle \Phi_1(x_1) \Phi_2 (x_2) \Phi_3(x_3) \rangle &= \delta^{(d)}(x_1-x_2) \frac{1}{(x_2-x_3)^{2\delta}}\frac{1}{(x_1-x_3)^{2\tilde{\delta}}} \cr
&= \delta^{(d)}(x_1-x_2) \frac{1}{(x_2-x_3)^{2\Delta_3}} \label{semil1}
\end{align}
where $\Delta_3 = \Delta_{1}+\Delta_2-d$. In particular, when $\Delta_1=d $ and $\Delta_3 = \Delta_2$, the two semi-local three-point functions are mutually compatible
\begin{align}
\langle \Phi_1(x_1) \Phi_2 (x_2) \Phi_3(x_3) \rangle = c_1 \delta^{(d)}(x_1-x_2) \frac{1}{(x_2-x_3)^{2\Delta_3}} + c_2 \delta^{(d)}(x_1-x_3) \frac{1}{(x_2-x_3)^{2\Delta_3}} \ , \label{semil}
\end{align}
which will become important later.

One may understand an origin of the semi-local terms from the effective action approach. Let us consider the following effective action 
\begin{align}
S_{\mathrm{eff}} = S_0 + \left( \int d^dx J^i(x)\Phi_i(x) + J^1(x) J^2(x) \Phi_3(x) \  \right) ,
\end{align}
the last term of which is conformal invariant if $\Delta_3 = \Delta_{1}+\Delta_2-d$, and reproduces \eqref{semil1}. A canonical example of such semi-local terms (appearing as the effective action) is the seagull term in scalar QED $A^\mu A_\mu |\phi|^2$, which is necessary to ensure the gauge invariance. 

We note that the semi-local term here is directly related to the local operator product expansions with delta functions studied in \cite{Schwimmer:2018hdl}. The semi-local term is equivalent to the operator product expansion
\begin{align}
\Phi_1(x) \Phi_2(y) = \delta^{(d)}(x-y) \Phi_3(x) + \cdots , 
\end{align}
which our results showed is conformal invariant for any $\Delta_1$ and $\Delta_2$ as long as $\Delta_3 = \Delta_1 + \Delta_2 -d$. 

The situation becomes more non-trivial with derivatives acting on the delta function. For example, contact three-point functions
\begin{align}
\langle \Phi_1(x_1) \Phi_2 (x_2) \Phi_3(x_3) \rangle = \delta^{(d)}(x_1-x_2)\partial_\mu \partial^\mu \delta^{(d)}(x_2-x_3) \ \label{3contd}
\end{align}
are conformal invariant only when $\Delta_2 = \frac{d}{2}+1$ and $\Delta_1 + \Delta_3 = \frac{3d}{2} + 1$, or when $\Delta_3 = \frac{d}{2}+1$ and $\Delta_1 + \Delta_2 = \frac{3d}{2} + 1$. 

Another interesting case is the semi-local three-point function
\begin{align}
\langle \Phi_1(X_1) \Phi_2 (X_2) J_M (X_3) \rangle = \partial^{X_1}_N \delta^{(d)}_k(X_1,X_2) \left(\eta^{NM}- \frac{X_3^NX_2^M}{(X_2X_3)}\right) (X_2X_3)^{-\tilde{\delta}} (X_1X_3)^{-\delta} 
\end{align}
with $\Delta_1 = 1+\delta$, $\Delta_2 = d+\tilde{\delta}$, $\Delta_3 = \delta + \tilde{\delta}$, which gives the conformal invariant semi-local terms
\begin{align}
\langle \Phi_1(x_1) \Phi_2 (x_2) J_\mu (x_3) \rangle \cr
= \partial_\nu \delta^{(d)}(x_1-x_2) \left(\eta^{\nu\mu}- 2\frac{(x_2^\nu - x_3^\nu)(x_2^\mu-x_3^\mu)}{(x_2-x_3)^2}\right) (x_2-x_3)^{-2\tilde{\delta}}(x_1-x_3)^{-2{\delta}} \label{confss}
\end{align}
in the Poincar\'e section. Note that although we would eventually set $x_1=x_2$, it is not allowed to set so in the last two terms of \eqref{confss}, which would give a different distribution.

Similarly, the semi-local three-point functions
\begin{align}
\langle \Phi_1(x_1) \Phi_2 (x_2) \Phi_3(x_3) \rangle = \partial^\mu \partial_\mu \delta^{(d)}(x_1-x_2) \frac{1}{(x_2-x_3)^{2\delta}} \frac{1}{(x_1-x_3)^{2\tilde{\delta}}}  \ \label{3semid}
\end{align}
are conformal invariant when $\Delta_1 +  \Delta_2 = d+2 + \Delta_3$ with $\delta + \tilde{d} = \Delta_3$.\footnote{It is curious to observe that only when $\Delta_1 = \frac{d}{2}+1$ (or $\Delta_2 = \frac{d}{2}+1$) one can construct the simple effective action $\int d^dx \left( \Phi_3 J^2 \partial_\mu \partial^\mu J^1 + J^3 \Phi_3 \right) $.} 
They correspond to the local operator product expansion
\begin{align}
\Phi_1(x_1) \Phi_2 (x_2)  =   \partial_\mu \delta^{(d)}(x_1-x_2) J_\mu(x_2) + \cdots \ , \label{Lapopv}
\end{align}
and
\begin{align}
\Phi_1(x_1) \Phi_2 (x_2)  =  \partial^\mu \partial_\mu \delta^{(d)}(x_1-x_2) \Phi_3(x_2) + \cdots \ . \label{LapOPE}
\end{align}

Let us now discuss possible physical origins of semi-local three-point functions. The most typical example comes from the Ward-Takahashi identity of a conserved current
\begin{align}
\langle \partial^\mu J_\mu(x_1) \Phi_1(x_2) \Phi_2(x_3) \rangle = \delta^{(d)}(x_1-x_2) \langle \delta \Phi_1(x_2) \Phi_2(x_3) \rangle + \delta^{(d)}(x_1-x_3) \langle \Phi_1(x_2) \delta \Phi_2(x_3) \rangle \ . 
\end{align}
Here the dimension of $\partial^\mu J_\mu$ is $d$ from the conservation, precisely in which case it behaves as a conformal primary. As we have discussed, it does not vanish only when $\Delta_1 = \Delta_2$ as dictated by the non-vanishing non-local conformal two-point functions.

The three-point contact terms also appear in anomalous correlation functions. As an example, let study the conformal invariance of the impossible anomaly in four dimensions:
\begin{align}
\langle \partial^\mu J_\mu(x_1) J_\rho(x_2) J_\sigma(x_3) \rangle = c (\partial^3_\rho \partial^2_\sigma -\delta_{\rho\sigma} \partial^2 \partial^3) \left( \delta^{(4)}(x_1-x_2) \delta^{(4)}(x_1-x_3) \right) \ ,
\end{align}
where the superscript indicates the coordinates we differentiate.
This anomalous conservation is conformal invariant in agreement with the fact that the impossible anomaly equation 
\begin{align}
D^\mu J_\mu = F_{\mu\nu} F^{\mu\nu}
\end{align}
is Weyl invariant. In our embedding space formalism, we see that $(\partial^3_\rho \partial^2_\sigma -\delta_{\rho\sigma} \partial^3\partial^2)$ acting on $ \delta^{(4)}_k(X_1,X_2) \delta_{\tilde{k}}^{(4)}(X_1,X_3)$ is conformal invariant only when $\delta^{(4)}_k(X_1,X_2)$ and $\delta^{(4)}_{\tilde{k}}(X_1,X_3)$ have projective weights $(2,2)$ (or $k=\tilde{k} = -2$).

This term is regarded as an impossible anomaly because the corresponding three-point function without the derivative is only semi-local%\footnote{Moreover, this three-point function is not conformal invariant while the anomaly is conformal invariant.}
\begin{align}
\langle  J_\mu(x_1) J_\rho(x_2) J_\sigma(x_3) \rangle = c (\partial^3_\rho \partial^2_\sigma -\delta_{\rho\sigma} \partial^2\partial^3)  (\Box^1 \partial^1_\mu \log(x_1- x_2)) \delta^{(4)}(x_1-x_3) + \text{perm} \ 
\end{align}
and unlike the conventional chiral anomaly
\begin{align}
D^\mu J_\mu = \tilde{F}_{\mu\nu} F^{\mu\nu}
\end{align}
there is no corresponding non-local three-point functions \cite{Costa:2011mg}\cite{Zhiboedov:2012bm}.  

The impossible anomalies have played an important role in the debate over whether the Pontryagin term is allowed on the trace anomaly in CP violating conformal field theories \cite{Nakayama:2012gu}\cite{Bonora:2014qla}\cite{Bonora:2017gzz}\cite{Bonora:2018obr}\cite{Bastianelli:2018osv}. The Pointryagin trace anomaly is related to the contact term \cite{Nakayama:2018dig}
\begin{align}
\langle T^{\mu}_{\mu}(x_1) T_{\sigma\rho}(x_2) T_{\alpha\beta}(x_3) \rangle = \epsilon_{\sigma \alpha \epsilon \kappa}[(\partial^2_\beta \partial^3_\rho - \partial^2\partial^3 \delta_{\beta\rho})(\partial^\epsilon(\delta^{(4)}(x_1-x_2)\partial^\kappa\delta^{(4)}(x_1-x_3))] + \mathrm{perm} \ .
\end{align}

Finally, the other applications of semi-local terms or semi-local operator product expansion appeared e.g. in \cite{Schwimmer:2018hdl}, which relates them to a non-trivial geometric structure of the conformal manifold. They argued that the semi-local terms in the operator product expansion
\begin{align}
\Phi_1(x) \Phi_2(y) = \delta^{(d)}(x-y) \Gamma_{12}^i \Phi_i(x) + \cdots
\end{align}
is related to the connection on the conformal manifold. They also showed concrete examples which contain the terms like \eqref{LapOPE} with derivatives on the delta functions.
We have showed that such semi-local operator product expansion are compatible with conformal invariance.

\section{Discussions}
In this paper, we have developed some techniques to study conformal invariant contact terms or semi-local terms. We have introduced representations of the delta function in  the embedding space formalism that can be used as a building block to construct such local correlation functions. As a complementary method, one may use the effective action approach to study the conformal invariance of these correlation functions. In both approaches, in order to address the conformal invariance of derivatives of delta functions, it is convenient to use the known facts about the representation theory of conformal algebra, but in principle one can reproduce these by using the embedding space method.

These conformal invariant contact terms or semi-local terms appear physically in the Ward-Takahashi identities. Here, we would like to remark that the Ward-Takahashi identity does not necessarily tell that the contact terms must be ``conformal invariant". To see this, let us consider the Ward-Takahashi identity for the translation invariance of two-point functions of scalar operators
\begin{align}
\langle \partial^\mu T_{\mu\nu} (x) \Phi_1(y) \Phi_2(z) \rangle = \delta^{(d)}(x-y) \langle \partial_\nu \Phi_1(y) \Phi_2(z) \rangle +  \delta^{(d)}(x-z) \langle  \Phi_1(y) \partial_\nu \Phi_2(z) \rangle \ . \label{WTT}
\end{align}
One would expect that since $\partial^\mu T_{\mu\nu}(x)$ transforms as a primary operator (because of the null vector condition), so would be the right hand side. However, each term of the right hand side is clearly not conformal invariant (in the sense that all three operators transform as primary operators) because of the derivative. This derivative is not something that we do not want, but it is precisely the term that generates the translation.

A resolution of the puzzle comes from the observation once we introduce the (space-time dependent) source term for $\Phi_i(x)$, the conservation of the energy-momentum tensor is lost (as $\partial^\mu T_{\mu\nu}(x) = (\partial_\nu J^i(x)) O_i(x)$), and accordingly although the compensated Weyl invariance may be preserved, the conformal symmetry, which relies on both the tracelessness and the conservation of the energy-momentum tensor, is not preserved. Of course, if we derive the Ward-Takahashi identity from the scratch, everything is manifest and \eqref{WTT} is the correct and consistent Ward-Takahashi identity. What we have learned here is simply that the meaning of ``conformal invariance" in contact terms or semi-local terms should require careful considerations. 

For a future direction, it would be an interesting question to study if the conformal invariance of local terms can be understood as an isometry of AdS in the holography. Also, generalizations of our formulation with spinors or supersymmetry must be of interest. The recent works have shown new anomalies in supersymmetry \cite{Papadimitriou:2017kzw}\cite{An:2017ihs}\cite{An:2019zok}\cite{Katsianis:2019hhg}\cite{Papadimitriou:2019gel}\cite{Papadimitriou:2019yug} and it will be interesting to recast these terms in our formalism.

\section*{Acknowledgements}
This work is in part supported by JSPS KAKENHI Grant Number 17K14301.

%%%%%%%%%%%%%%%%%%%%%%%%%%%%%%%%%%%%%%%%%%%%%%%%%%%%%%%%%%%%%%%%%%%%%%%%%%%%%%%%%%

\appendix
\section{A check of special conformal invariance in coordinate space}
In this appendix, we  perform a check of the special conformal invariance of some two-point functions directly in the coordinate space for $d=1$. The generalization to higher dimension should be straightforward.

Let us begin with the usual non-local two-point functions 
\begin{align}
\langle O_{\Delta_1}(x_1) O_{\Delta_2}(x_2) \rangle = \frac{1}{(x_1-x_2)^{\Delta_1 + \Delta_2}} \ .
\end{align} 
Assuming that $O_{\Delta_1}$ and $O_{\Delta_2}$ are primary operators, the left hand side transforms under the special conformal transformation $x \to x + \epsilon x^2$ as
\begin{align}
(1+2\epsilon x_1)^{-\Delta_1} (1+2\epsilon x_2)^{-\Delta_2} \langle O_{\Delta_1}(x_1) O_{\Delta_2}(x_2) \rangle \ .
\end{align}
On the other hand, the right hand side transforms as
\begin{align}
\frac{1}{(x_1 + \epsilon x_1^2 - x_2 - \epsilon x_2^2)^{\Delta_1 + \Delta_2}} = \frac{1-\epsilon(x_1 + x_2)(\Delta_1 + \Delta_2)}{(x_1 - x_2)^{\Delta_1 + \Delta_2}} \ .
\end{align} 
The equality holds only if $\Delta_1 = \Delta_2$, which is the well-known constraint on the two-point functions.

The constraint is weaker in the contact term without derivatives on the delta function. Let us consider 
\begin{align}
\langle O_{\Delta_1}(x_1) O_{\Delta_2}(x_2) \rangle = \delta(x_1-x_2) 
\end{align} 
with $\Delta_1 + \Delta_2 = 1$ from the scale invariance.
Assuming that $O_{\Delta_1}$ and $O_{\Delta_2}$ are primary operators, the left hand side transforms under the special conformal transformation $x \to x + \epsilon x^2$ as
\begin{align}
(1+2\epsilon x_1)^{-\Delta_1} (1+2\epsilon x_2)^{-\Delta_2} \langle O_{\Delta_1}(x_1) O_{\Delta_2}(x_2) \rangle \cr
= ( 1-2\epsilon x_1 (\Delta_1 + \Delta_2)) \delta(x_1 - x_2) \ . \label{a1}
\end{align}
On the other hand, the right hand side transforms as
\begin{align}
\delta(x_1 + \epsilon x_1^2 - x_2 - \epsilon x_2^2) \cr
= (1-2\epsilon x_1) \delta(x_1 -x_2) \ . \label{a2}
\end{align} 
The equality between \eqref{a1} and \eqref{a2} holds as long as $\Delta_1 + \Delta_2 = 1$ and there is no further constraint such as $\Delta_1 = \Delta_2$. 

Let us finally consider the contact term with a derivative acting on the delta function.
\begin{align}
\langle O_{\Delta_1}(x_1) O_{\Delta_2}(x_2) \rangle = \partial_{x_1} \delta(x_1-x_2) 
\end{align} 
with $\Delta_1 + \Delta_2 = 2$ from the scale invariance. Assuming that $O_{\Delta_1}$ and $O_{\Delta_2}$ are primary operators, the left hand side transforms under the special conformal transformation $x \to x + \epsilon x^2$ as
\begin{align}
(1+2\epsilon x_1)^{-\Delta_1} (1+2\epsilon x_2)^{-\Delta_2} \langle O_{\Delta_1}(x_1) O_{\Delta_2}(x_2) \rangle \cr
= (1+2\epsilon x_1)^{-\Delta_1} (1+2\epsilon x_2)^{-\Delta_2} \partial_{x_1} \delta(x_1 - x_2) \ . \label{c1}
\end{align}
Note that unlike the case above, we cannot set $x_1=x_2$ in front of the derivative of the delta function.\footnote{For example, $x\partial_x \delta(x-y)$ is different from $y\partial_x \delta(x-y)$ as a distribution, which we can easily see by multiplying $f(x)$ and integrating it over $x$.} 
On the other hand, the right hand side transforms as
\begin{align}
(1-2\epsilon x_1) \partial_{x_1} \delta(x_1 + \epsilon x_1^2 - x_2 - \epsilon x_2^2) \cr
= (1-2\epsilon x_1) \partial_{x_1}((1-2\epsilon x_1) \delta(x_1 -x_2)) \ . \label{c2}
\end{align} 
The equality between \eqref{c1} and \eqref{c2} holds only if $\Delta_1 = 1$ and $\Delta_2=1$. 
This is consistent with our embedding space formalism.

\end{document}